# Towards Unified Neural Decoding with Brain Functional Network Modeling

Di Wu [1,2,4*], Linghao Bu [3*], Yifei Jia [1,2,4*], Lu Cao [2], Siyuan Li [2], Siyu Chen [4], Yueqian Zhou [2,4,5], Sheng Fan [4], Wenjie Ren [2,4,5], Dengchang Wu [6], Kang Wang [6], Yue Zhang [1,2,4#], Yuehui Ma [3#], Jie Yang [1,2,4,5#], Mohamad Sawan [1,2,4,5#]

[1]Center of Excellence in Biomedical Research on Advanced Integrated-on-chips Neurotechnologies (CenBRAIN), School of Engineering, Westlake University, Hangzhou, China
[2]School of Engineering, Westlake University, Hangzhou, Zhejiang Province, China
[3]Epilepsy Center, Department of Neurosurgery, Zhejiang University School of Medicine First Affiliated Hospital, Hangzhou, Zhejiang Province, China
[4]Integrated-on-Chips Brain-Computer Interfaces Zhejiang Engineering Research Center, Westlake University, Hangzhou, China
[5]Westlake Institute for Optoelectronics, Hangzhou, Zhejiang, China.
[6]Epilepsy Center, Department of Neurology, Zhejiang University School of Medicine First Affiliated Hospital, Hangzhou, Zhejiang Province, China

[*] These authors contributed equally to the work.
[#]Corresponding Author.

# Abstract

Recent achievements in implantable brain-computer interfaces (iBCIs) have demonstrated the potential to decode cognitive and motor behaviors with intracranial brain recordings; however, individual physiological and electrode implantation heterogeneities have constrained current approaches to neural decoding within single individuals, rendering interindividual neural decoding elusive. Here, we present Multi-individual Brain Region-Aggregated Network (MIBRAIN), a neural decoding framework that constructs a whole functional brain network model by integrating intracranial neurophysiological recordings across multiple individuals. MIBRAIN leverages self-supervised learning to derive generalized neural prototypes and supports group-level analysis of brain-region interactions and inter-subject neural synchrony. To validate our framework, we recorded stereoelectroencephalography (sEEG) signals from a cohort of individuals performing Mandarin syllable articulation. Both real-time online and offline decoding experiments demonstrated significant improvements in both audible and silent articulation decoding, enhanced decoding accuracy with increased multi-subject data integration, and effective generalization to unseen subjects. Furthermore, neural predictions for regions without direct electrode coverage were validated against authentic neural data. Overall, this framework paves the way for robust neural decoding across individuals and offers insights for practical clinical applications.

# Main

The intersection of deep learning techniques with iBCIs has significantly advanced neurorehabilitation, facilitating the restoration of sensory, motor, and communication functions in individuals with neurological impairments[1-5]. Over recent decades, substantial progress has been made, notably in increasing decoding accuracy and longevity. However, most existing neural decoding frameworks within iBCIs remain highly individualized, requiring extensive calibration for each patient[6-9], thus limiting their practical deployment to controlled laboratory settings, and hindering broader clinical adoption[10].

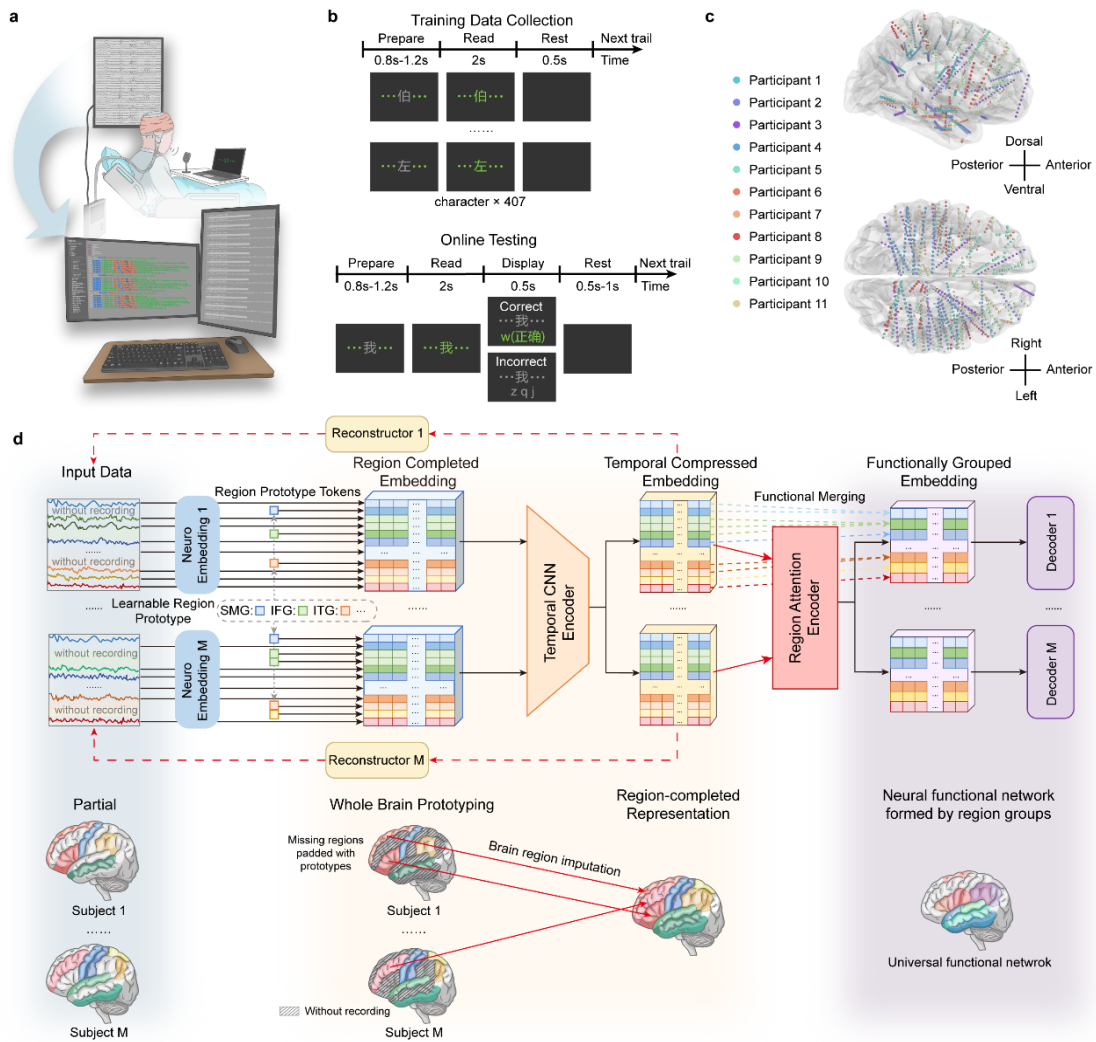

**Fig. 1. | A universal neural decoding framework verified on the task of phoneme articulation. a**, Participants implanted with stereoelectroencephalography (sEEG) electrodes performed tasks involving reading Mandarin Chinese characters aloud or silently in response to visual cues. Recorded neural signals were transmitted to a central server and used to train the proposed MIBRAIN model to decode and predict the initial consonants of articulated characters. Following the training phase, participants' neural activity was streamed in real-time to the server for immediate neural decoding of phoneme articulation. **b**, A visual prompting paradigm was developed for both training data collection and real-time online decoding experiments. Participants were instructed to audibly or silently articulate Mandarin Chinese characters representing 407 distinct syllables, covering all possible combinations of initial and final phonemes. During the online decoding phase, decoded predictions were displayed to participants in real-time to provide immediate visual feedback. **c**, Implantation sites of selected electrodes from all participants are illustrated on the standard Montreal Neurological Institute (MNI) template brain. Electrodes from each participant are depicted using distinct colors to clearly differentiate implantation sites. Anatomical orientation indicated by directional labels ensures clarity. **d**, Overview of the MIBRAIN framework. Upper panel illustrates the detailed algorithmic architecture of the MIBRAIN framework, with corresponding functional brain network modeling depicted conceptually in the lower panel. Matched colors across panels indicate corresponding sections between algorithmic modules and their related

neural functional representations. During the whole-brain functional prototyping stage, MIBRAIN imputes representations for brain regions lacking direct recordings by leveraging prototypes derived from other regions or participants, thus constructing a complete-region neural representation for each subject. Subsequently, in the neural decoding stage, MIBRAIN leverages collaborative interactions between brain regions to identify groups forming functionally connected regions essential for accurate decoding.

A promising strategy to address these limitations is to integrate neurophysiological data across multiple subjects to establish a robust, cross-subject neural decoding framework. Although group-level neuroimaging studies have consistently demonstrated neural commonalities among individuals[16-18], effectively combining intracranial signals across diverse brain regions and subjects remains challenging. Over the past decade, various computational approaches have been proposed, including instance-weighting[19,20], feature transformation[21,22], and adversarial learning[23,24]. These methods typically assume homogeneous experimental conditions, such as standardized electrode placements in EEG systems (e.g., 10-20 or 10-10 systems). In contrast, intracranial recordings present significant heterogeneity in electrode placements and counts across subjects, underscoring a critical need for advanced computational frameworks capable of handling this variability. Recently, novel deep learning approaches have demonstrated effectiveness in processing EEG signals with variable channel counts and temporal dimensions by converting them into channel-wise temporal token sequences[25,26]. However, these approaches mainly address data dimensionality without fully leveraging the underlying neurophysiological mechanisms and brain network interactions from a neuroscience perspective.

Complex cognitive functions, such as language processing, arise from dynamic interactions across spatially distributed but functionally interconnected brain regions. Despite the acknowledged importance of these neural connectomes and networks, contemporary decoding approaches utilizing intracranial recordings primarily depend on data from single subjects, often collected via electrode arrays targeting specific cortical regions (e.g., ventral sensorimotor cortex for speech prostheses)[2,4,9,11]. Emerging studies suggest that incorporating signals from broader brain regions could enhance both the accuracy and universality of neural decoding in iBCIs[12-15]. Nonetheless, efforts in this direction are constrained by the limited electrode coverage achievable within individual subjects, given the invasiveness of current intracranial electrode technology.

Addressing these critical gaps, we propose the Multi-individual Brain Region-Aggregated Network (MIBRAIN), an advanced modeling framework designed to integrate intracranial recordings from multiple subjects with distinct electrode implantation schemes to achieve efficient, generalized neural decoding. MIBRAIN aggregates brain recordings across multiple individuals, utilizing a self-supervised learning approach to derive representative region prototypes capturing generalized neural characteristics under identical neural activity conditions. For each participant, these prototypes enable inference of neural activity even in regions lacking direct recordings, thus constructing a comprehensive functional brain network model. Through a novel temporal and regional fusion mechanism guided by neurological annotations, MIBRAIN aligns neural representations from different subjects into a unified neural representation space, facilitating group-level analyses of collaborative brain-region interactions and inter-subject neural synchrony.

We demonstrate the efficacy of MIBRAIN in real application through a study on Mandarin Chinese phoneme articulation involving 11 participants, evaluated using real-time online decoding experiments and systematic offline analyses. Our results indicate that MIBRAIN significantly enhances decoding performance for both audible and silent speech by integrating multi-subject data. Additionally, we observed a consistent improvement in decoding accuracy with the progressive inclusion of more participants. Crucially, we validated the correlation of neural features predicted in brain regions lacking electrode coverage against authentic neural representations and confirmed MIBRAIN's ability to generalize effectively to previously unseen subjects. Finally, our framework robustly identifies collaborative inter-region dynamics throughout phoneme articulation processes, underscoring MIBRAIN's potential applicability for patients with severe neurological impairments, where ideal electrode placements or extensive data collection are impractical.

## Results

### Overview of MIBRAIN

We present MIBRAIN, a whole functional network modeling framework designed to perform neurological decoding across multiple individuals. As illustrated in Fig. 1d, MIBRAIN comprises two main stages: whole functional network prototyping and neural decoding.

During the whole functional network prototyping stage, we first define a comprehensive set of brain regions relevant to a specific neural activity, an approach we term whole functional network modeling. MIBRAIN generates region-wise neural tokens from each subject's intracranial recordings using dedicated neural embedding layers, which fuse information across channels within a brain region while accounting for differences in electrode implantation. For a certain subject, some brain regions may lack recordings due to non-implantation, and thus, the resulting neural tokens represent only a partial view of the functional network. To complete the representation, these partial tokens are concatenated with learnable region prototype tokens corresponding to the regions with no recordings, forming a region-complete neural representation. This complete neural representation is then passed through a temporal CNN for unified feature extraction. Moreover, MIBRAIN employs a masked autoencoding strategy that randomly masks out a subset of regional tokens and reconstructs the missing information, thereby optimizing each prototype token by propagating information from other brain regions within the same subject or from the same brain region across multiple subjects.

In the neural decoding phase, the pre-trained framework, consisting of the neural embedding layers, region prototype tokens, and the temporal CNN encoder, is further augmented with a region attention encoder and MLP-based prediction heads (decoders), both initialized with random weights. The region attention encoder is designed to capture the collaborative relationships among brain regions during different stages of the neural activity, enabling enhanced information fusion across region groups performing the same function. The entire framework is then fine-tuned using supervised labels for each recording segment, ensuring that the extracted neural features are robustly aligned with the underlying decoding task.

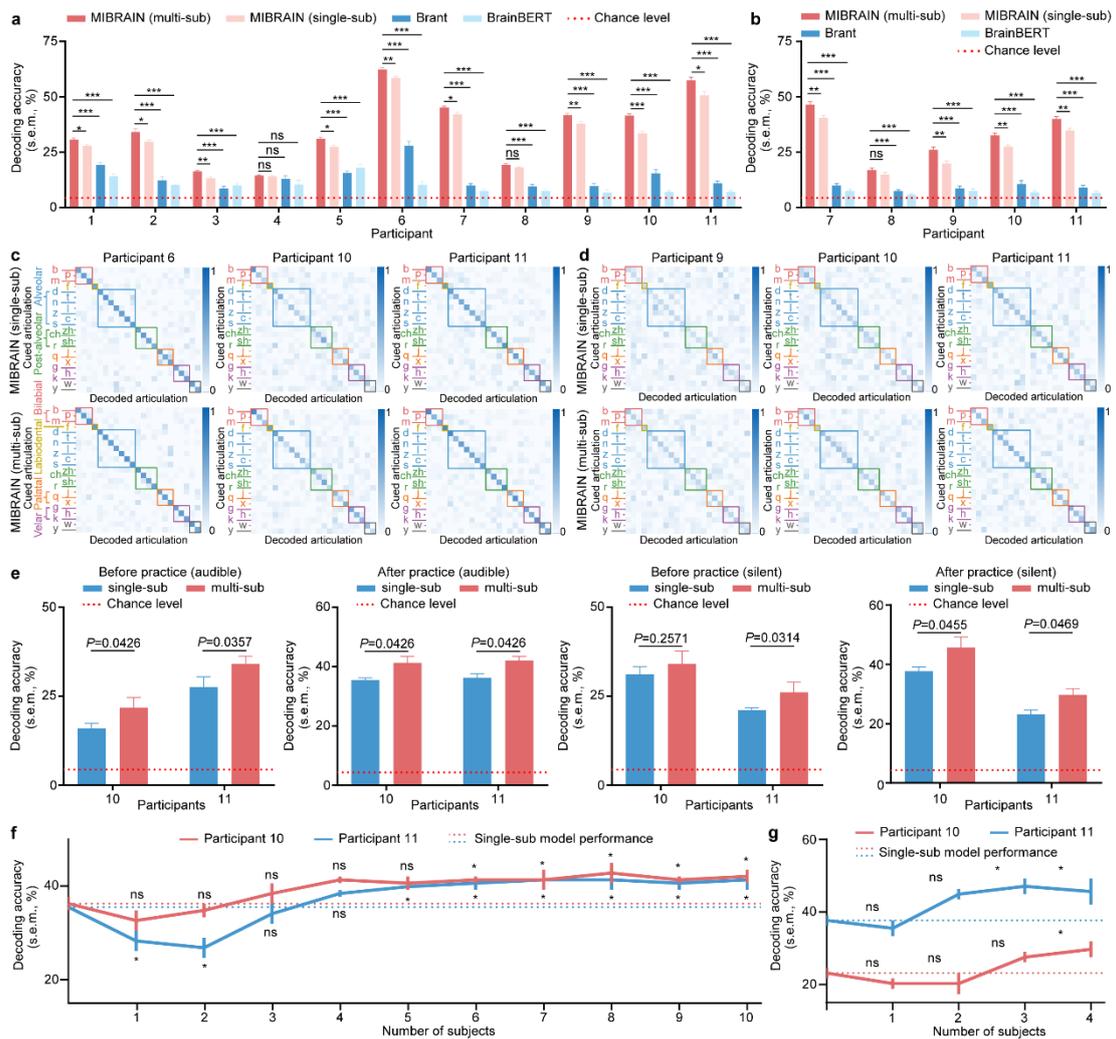

**Fig. 2. | Decoding phoneme articulation performance using MIBRAIN. a**, Offline decoding accuracy (mean ± s.e.m.) of MIBRAIN in audible articulation measured across 11 participants, compared to Brant, BrainBERT, and a chance-level baseline. MIBRAIN (multi-sub) and MIBRAIN (single-sub) refer to models trained with data aggregated across all participants and individually-trained models, respectively. **b**, Offline decoding accuracy of MIBRAIN in silent articulation measured in five participants, compared to alternative decoding approaches. **c**, Audible phoneme decoding accuracy displayed by consonant initials grouped according to place of articulation, comparing MIBRAIN (single-sub, top row) and MIBRAIN (multi-sub, bottom row) performance for participants 6, 10, and 11. **d**, Offline silent phoneme

decoding accuracy by consonant initials using both MIBRAIN variants (single-sub and multi-sub) for participants 9, 10, and 11. **e**, Online decoding accuracy of MIBRAIN for audible and silent articulation, measured before and after one hour of free practice for participants 10 and 11. Significance comparison is conducted using paired t-tests on the decoding performance for each initial consonant. **f**, Online decoding performance of MIBRAIN as a function of the number of participants' data included in model training for audible articulation decoding, for participants 10 and 11. **g**, Online decoding accuracy of MIBRAIN in silent articulation decoding as a function of the number of participants' data used in training, for participants 10 and 11.

**MIBRAIN enables generalized neural decoding**

To comprehensively demonstrate the decoding capabilities of MIBRAIN, we conducted a proof-of-concept experiment evaluating phoneme articulation decoding performance. Intracranial recordings were acquired from 11 subjects instructed to articulate Mandarin Chinese characters, either audibly or silently (mimed). Considering Mandarin Chinese encompasses a vocabulary exceeding 50,000 characters, participants read a specially designed set of 407 monosyllabic characters representing all possible combinations of initial and final phonemes. For simplicity and generalizability, we specifically focused on decoding the initial phoneme elements (23 distinct initials in total) articulated by the subjects.

We first demonstrate that integrating intracranial recordings from multiple subjects enables MIBRAIN to robustly capture neural functional networks underlying phoneme articulation, leading to significantly improved decoding performance. To rigorously assess the ability of MIBRAIN to generalize neural decoding capabilities across participants, we compared its decoding accuracy against two baseline approaches: Brant[27], BrainBERT[28] and MIBRAIN variants trained solely on individual subjects. The baseline models utilize large-scale neural recordings and employ extensive pre-training strategies; however, they still require participant-specific fine-tuning and thus fail to achieve strong cross-participant decoding generalization. Moreover, these existing approaches have not fully leveraged multi-subject data to elucidate functional collaborations across brain regions, resulting in decoding accuracy barely exceeding chance levels. By contrast, both single-subject MIBRAIN (MIBRAIN-single-sub) and multi-subject MIBRAIN (MIBRAIN-multi-sub) substantially outperformed the baseline approaches and exceeded chance-level decoding performance on both audible and silent articulation tasks for all participants except participant four (all comparisons P < 0.001; Fig. 2a and Fig. 2b). Crucially, MIBRAIN multi-sub displayed consistently superior decoding accuracy compared to its single-subject counterpart for most participants. Specifically, average decoding accuracy improvements achieved by MIBRAIN multi-sub over MIBRAIN single-sub were 8.08% and 6.83% for audible articulation, and 5.10% and 4.97% for silent articulation, for participants 10 and 11 respectively. We observed notably lower decoding performance for participants 3, 4, and 8 compared to other participants. Although these three participants are fluent native speakers of Mandarin Chinese, their distinctive regional dialectal pronunciation significantly influenced their phoneme articulation patterns. For example, they frequently confused certain initial consonants such as 'n' and 'l', and experienced difficulty distinguishing places of articulation, specifically alveolar consonants 's' and retroflex consonants 'sh'. Consequently, these pronunciation features decreased discriminability among their initial consonant articulations at the neural level, ultimately resulting in lower overall decoding accuracy.

The articulation of Mandarin initials can be characterized by the place of articulation (POA): bilabial, labiodental, alveolar, retroflex, alveolo-palatal and velar). We explored how the additional multi-subject training afforded by MIBRAIN multi-sub improved decoding performance at the initial consonant level from the perspective of POA. During audible articulation decoding, for example, the confusion matrix from participant 6 demonstrated clearer discrimination among alveolo-palatal initials ('j', 'q', 'x') with MIBRAIN multi-sub, whereas these consonants were frequently misclassified using the single-subject variant (Fig. 2c). Similarly, during silent articulation decoding, MIBRAIN multi-sub consistently enhanced the distinction among velar initials, typically challenging to decode due to minimal overt articulation, leading to improved decoding accuracy in participants 9 and 10 (Fig. 2d).

**Real-time decoding of MIBRAIN**

We conducted real-time decoding experiment with participant 10 and 11. Intracranial signals recorded during character articulation were streamed live to a server for real-time decoding of the articulated initial phoneme (Fig. 1a). We compared decoding performance between the model trained on aggregated data from all participants (MIBRAIN-multi-sub) and models trained individually per participant (MIBRAIN-single-sub) for both audible and silent articulation modes (Fig. 2e). In this procedure, streamed neural recordings were simultaneously provided to both models for online decoding, allowing parallel performance assessment. Each participant completed two randomized trials per articulation mode, separated by 10-minute rest intervals. Subsequently, participants engaged in a one-hour free-practice session allowing for self-paced decoding exercises and personalized rest periods. We observed that online decoding accuracies for both models and articulation modes were significantly above chance level even before the practice session, despite the absence of any subject-specific calibration or training. Importantly, a consistent increase in decoding accuracy was observed following the practice period across both articulation modes, likely attributable to participants becoming increasingly accustomed to the decoding system and stabilizing their articulation patterns. Consequently, we consider the post-practice performance a more accurate representation of MIBRAIN's decoding capability. Notably, MIBRAIN (multi-sub) consistently outperformed MIBRAIN (single-sub) across all conditions. Specifically, following the practice sessions, average decoding accuracy improvements for MIBRAIN (multi-sub) compared to MIBRAIN (single-sub) were 3.62%, 2.90%, 7.97%, and 4.35% for participants 10 and 11 in audible and silent articulation modes, respectively.

**Integrating more participants' data improves decoding**

Next, we investigated how decoding performance of MIBRAIN in real-time articulation prediction changes as more participants' data are included into model training. Specifically, we evaluated performance for participants 10 and 11 in both audible and silent articulation online decoding tasks, comparing a reference model trained solely on each participant's own data (MIBRAIN-single-sub) against models incrementally augmented by adding data from additional subjects (Fig. 2f, g). As participant order can impact model outcomes and exhaustive enumeration of participant orderings is impractical, we defined our participant inclusion sequence based on the order in which participants enrolled in the study. Thus, for each of participants 10 and 11, a total of eleven models were trained, incrementally incorporating training data from between 1 (single-subject baseline) and all 11 participants. During the online decoding sessions, intracranial recordings were livestreamed to all trained models in parallel to precisely assess the influence of scaling participant data on decoding accuracy.

Overall, decoding accuracy exhibited a characteristic trend as training set size increased: initial performance degradation, followed by gradual improvement (Fig. 2f, g). Notably, integrating data from a small cohort (fewer than three additional participants) initially led to a clear drop in decoding accuracy, for instance, participant 11 demonstrated a significant performance reduction during audible decoding when data from fewer than three participants were added. This early performance decline likely arises due to inter-subject heterogeneity in neural characteristics and electrode implantation sites, which MIBRAIN initially struggles to reconcile when trained on limited subject data. Subsequently, with the addition of more participant data, we observed consistent improvements in decoding accuracy. This trend suggests that MIBRAIN progressively aligns neural representations from distinct individuals into a coherent, unified neural representational space, thereby enhancing decoding performance. Specifically, decoding accuracy improvements became statistically significant ($P < 0.05$) once data from at least six additional participants were incorporated into training for both participants in audible articulation. Given that fewer participants completed the silent articulation task, our conclusions regarding performance scalability with increasing data remain preliminary. Still, participant 10 showed continuing upward trends, suggesting potential for further performance enhancement as additional participant data becomes available.

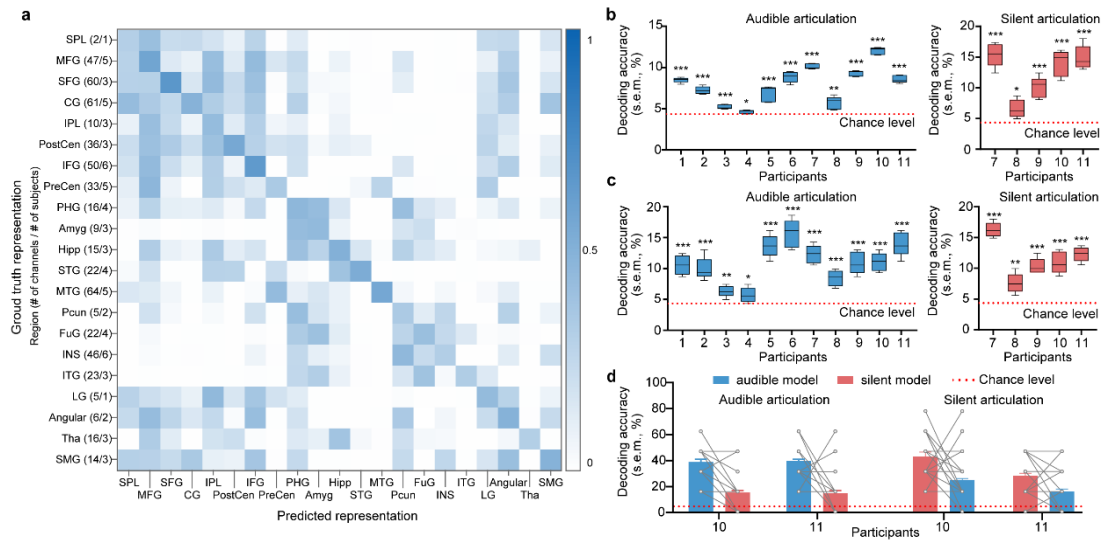

**Fig. 3. | Verification and generalization ability of MIBRAIN. a**, Correlation matrices comparing the removed and imputed neural representations against the ground-truth neural representation obtained from MIBRAIN (multi-sub) for each brain region involved in the study. For each region, the number of participants contributing data and total number of recorded channels are indicated. Brain-region abbreviations are as follows: SPL, superior parietal lobule; MFG, middle frontal gyrus; SFG, superior frontal gyrus; CG, cingulate gyrus; IPL, inferior parietal lobe; PostCen, postcentral gyrus; IFG, inferior frontal gyrus; PreCen, precentral gyrus; PHG, parahippocampal gyrus; Amyg, amygdala; Hipp, hippocampus; STG, superior temporal gyrus; MTG, middle temporal gyrus; Pcun, precuneus; FuG, fusiform gyrus; INS, insula; ITG, inferior temporal gyrus; LG, lingual gyrus; Angular, angular gyrus; Tha, thalamus; SMG, supramarginal gyrus. **b**, Offline decoding accuracies achieved using predicted neural representations (from MIBRAIN-multi-sub) for brain regions without electrode implantation, shown individually for each participant during audible and silent articulation tasks. **c**, Offline decoding accuracies of MIBRAIN for unseen subjects, calculated by training the model on data from all participants except the tested individual, across both audible and silent articulation modes. **d**, Online phoneme decoding accuracies for participants 10 and 11, comparing models separately trained using audible (audible model) or silent (silent model) articulation data, evaluated during real-time audible and silent articulation decoding tasks. Individual grey dots denote decoding performance for each initial consonant.

## MIBRAIN generalizes to unseen subjects

We further validated whether MIBRAIN effectively modeled the functional neural network in a subject-agnostic manner, thus capturing inter-subject neural synchrony. To assess this, we conducted a decoding experiment using a leave-one-subject-out cross-validation approach for both audible and silent articulation tasks: Specifically, we trained MIBRAIN models on neural data aggregated from all participants except one (the held-out test subject), and we then evaluated decoding performance using data from this unseen subject. Although decoding accuracy was lower than that of models trained on the complete dataset (i.e., including the test subject's own data), performance remained significantly above the chance baseline, demonstrating robust inter-subject generalization capabilities (Fig. 3c). Interestingly, the decoding performance for the silent articulation task on unseen subjects was closer to the accuracy levels obtained from models trained with all participants' data than was observed for the audible articulation task. We hypothesize that this improved generalization in the silent articulation condition may result from more exaggerated and stereotyped mouth movement patterns and thus homogeneous neural activity across different individuals during silent articulation.

**Cross-decoding between audible and silent articulation**

In this work, MIBRAIN constructed distinct decoding models associated with audible and silent phoneme articulation processes. Here we assessed the cross-task decoding performance of MIBRAIN by investigating the extent to which the network model built from audible articulation data ("audible model") could generalize to decoding silent articulation, and vice versa, by conducting real-time cross-decoding experiments with participants 10 and 11. Specifically, neural recordings acquired during both audible and silent articulation tasks were decoded using both the audible and silent models (based on MIBRAIN-multi-sub). While cross-decoding (audible model decoding silent articulation, and silent model decoding audible articulation) yielded lower performance compared to decoding by task-specific models, the decoding accuracies were still significantly higher than the baseline (Fig. 3d).

Interestingly, decoding audible articulation using the silent model exhibited a larger decrease in accuracy relative to decoding silent articulation using the audible model. We hypothesize this asymmetry arises because silent articulation lacks auditory feedback processing, thus, the silent articulation network modeled by MIBRAIN does not capture neural activity associated with auditory regions. In contrast, the audible articulation network inherently encompasses neural processes associated with silent articulation, thereby enabling comparatively better generalization and decoding performance in cross-task testing.

**Imputed brain regions correlate with ground-truth representations**

MIBRAIN integrates neural data across multiple participants, each contributing partial neural recordings from a subset of the targeted brain regions. Because electrode implantation is limited to specific regions in each individual, no single participant possesses a complete neural recording across all regions of interest. To overcome this, MIBRAIN first generates prototype representations for brain regions without direct recordings during the whole-network prototyping stage, and subsequently aggregates related neural information from available recordings in other regions or subjects. These aggregated representations thus provide an "imputed" neural activity pattern for unrecorded brain regions, enabling comprehensive neural decoding. We next verified whether the neural representations predicted for these imputed regions aligned closely with actual recorded neural representations. To perform this validation, we considered the representations derived by MIBRAIN-multi-sub for directly recorded regions as the "ground truth". We systematically removed one recorded region at a time from each participant's data, obtained predicted representations for that region, and then correlated each predicted representation with the ground-truth representation from the corresponding region and all other regions. We observed strong similarity between predicted and actual neural representations for each removed region, as indicated by high correlation values along the diagonal of the correlation matrix (Fig. 3a).

Furthermore, we found a significant correlation between predicted representation of Broca's area (i.e. IFG) and Wernicke's area (i.e. SMG and Angular), consistent with the cortical termination of arcuate fasciculus. These correlations were also identified within limbic system, such as amygdala, hippocampus, and regions in limbic lobe. The identification of these correlations supports the "authenticity" of the imputed regional representations, and further demonstrated the ability of MIBRAIN to capture the underlying functional collaboration between brain regions.

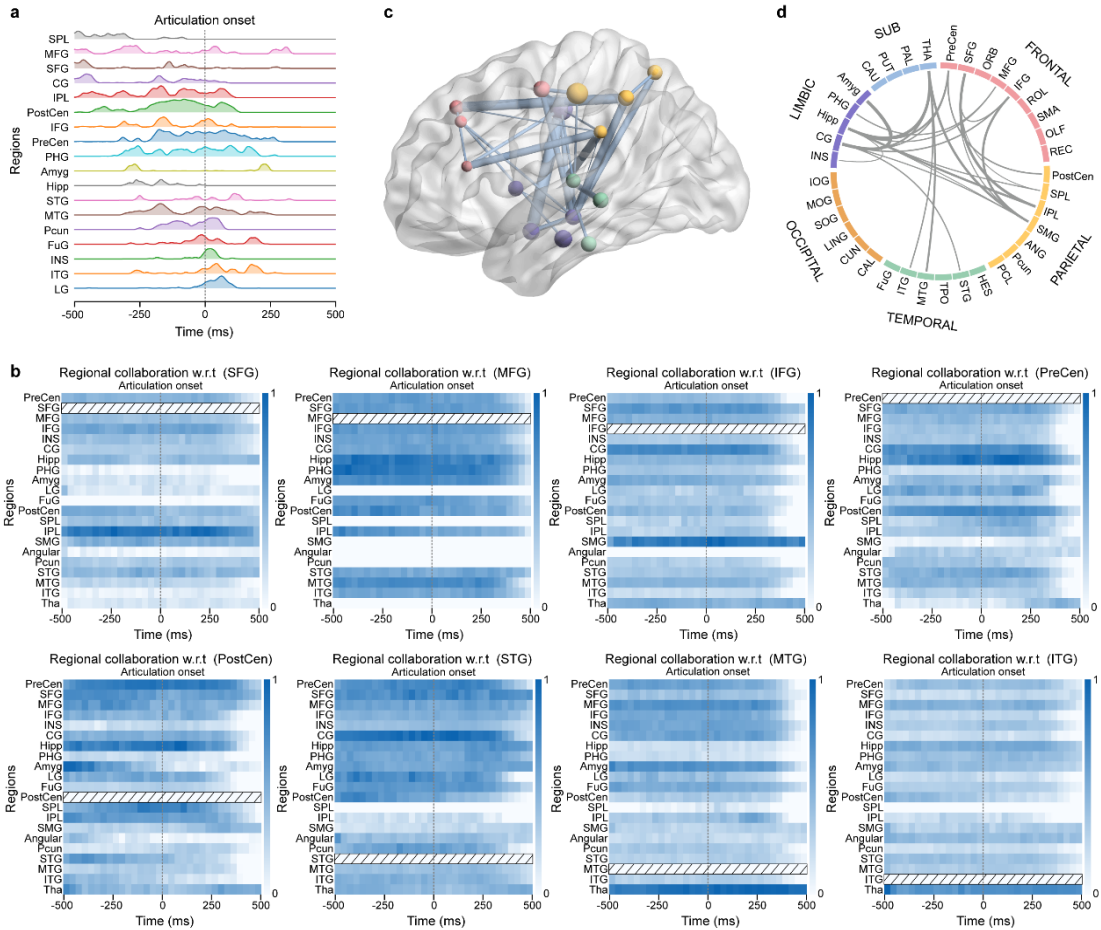

**Fig. 4. | Anatomical contributions and collaborations during phoneme articulation. a**, The contribution of each anatomical area is quantified by computing the gradient of the loss function with respect to neural recordings (for methodological details, see "Calculation of brain region contribution"). Shaded areas represent kernel density estimations of these contributions, averaged across all data samples and participants. The grey dotted line indicates the onset of articulation. Brain region abbreviations are as follows: SPL, superior parietal lobule; MFG, middle frontal gyrus; SFG, superior frontal gyrus; CG, cingulate gyrus; IPL, inferior parietal lobe; PostCen, postcentral gyrus; IFG, inferior frontal gyrus; PreCen, precentral gyrus; PHG, parahippocampal gyrus; Amyg, amygdala; Hipp, hippocampus; STG, superior temporal gyrus; MTG, middle temporal gyrus; Pcun, precuneus; FuG, fusiform gyrus; INS, insula; ITG, inferior temporal gyrus; LG, lingual gyrus; Angular, angular gyrus; Tha, thalamus; SMG, supramarginal gyrus. **b**, Functional collaboration of partial brain regions with respect to a target region (indicated by rectangles filled with diagonal lines). Collaboration strength was quantified based on the frequency at which each region was grouped with the target region across samples and participants (see "Calculation of Brain Region Collaboration" for methodological details). Darker colors represent greater likelihood that a region forms a functional group with the target brain region. We provide the functional collaboration for the remaining brain regions in Fig. S5. **c**, Functional collaboration among brain regions averaged across all target areas. The diameter of each sphere represents the averaged contribution of the corresponding brain region over the entire temporal domain. Lines connecting regions indicate collaboration strength, with thicker lines denoting stronger collaboration. Distinct colors represent brain regions located in different lobes. **d**, Functional collaboration of different brain regions with thicker lines denoting stronger collaboration.

**Neural decoding with imputed regional representations**

To further verify that the neural representations imputed by MIBRAIN indeed reflect meaningful neural activity relevant to the phoneme articulation decoding tasks, we performed decoding analyses using solely these imputed brain-region representations. We evaluated decoding accuracy separately for audible and silent articulations. The decoding performance significantly exceeded chance level by a wide margin ($P < 0.001$; Fig. 3b, left panel) in all participants except participants 4 and 8 during audible decoding. Similarly, decoding accuracy in the silent articulation task showed a clear improvement above the chance baseline (Fig. 3b, right panel). These results confirm our hypothesis that MIBRAIN effectively propagates task-relevant neural information, either from other recorded brain regions within the same participant or from homologous regions across multiple participants, to produce informative representations for regions lacking direct electrode implantation.

**Dynamic contribution of brain regions**

To investigate the dynamic involvement of brain region during speech production, we computed a region-specific contribution score during the whole articulation time course. In particular, we focused on the audible articulation task, for which articulation onset timings could be precisely identified by expert linguistic annotation based on audio recordings. We visualized the regional contributions from 500 milliseconds before to 500 milliseconds after the determined articulation onset.

As illustrated in Fig. 4a, the superior frontal gyrus (SFG) exhibited the earliest significant contribution in the time scale, suggesting its critical role in initiating speech. Following SFG activation, the middle frontal gyrus (MFG) displayed considerable engagement prior to speech onset. Subsequently, pronounced contributions emerged in the precentral and postcentral gyrus, reflecting their direct involvement in muscle control necessary for speech articulation. Additionally, the inferior frontal gyrus (IFG) showed substantial activity both preceding and during speech onset, consistent with its established role in semantic and phonological processing. Post articulation onset, notable activity shifted towards the superior temporal gyrus (STG), aligning with its function in auditory feedback processing. Beyond these traditionally recognized perisylvian language regions, notable contributions were also detected in various cortical areas such as the parahippocampal gyrus, cingulate gyrus, and inferior temporal gyrus, as well as subcortical structures including the hippocampus, amygdala, and thalamus. The widespread involvement of these additional regions underscores the extensive and dynamic collaboration between language-specific and broader cognitive networks during speech production.

**Collaborative brain-region interactions**

We further investigated the dynamic functional correlation between brain regions that showed high contribution in the phoneme articulation decoding explored by MIBRAIN. A correlated neural activity was found between IFG and SFG, as well as SMG, corresponding the cortical termination of two main language-related white matter tracts, frontal aslant tract and arcuate fasciculus, respectively (Fig. 4b). More interestingly, a wide collaboration between hippocampus and regions in language network (e.g. PreCen, PostCen and SMG) and other cognitive networks (e.g. cingulate gyrus, thalamus, and insula) (Fig. 4c, d). This finding reiterated the complexity of the neural mechanism of speech production and MIBRAIN could effectively capture the complex neural collaboration across the whole brain.

**Discussion**

In this study, we introduced MIBRAIN, a unified neural decoding framework designed to address key challenges inherent in multi-subject neural decoding tasks. Specifically, the proposed framework effectively manages inter-subject data heterogeneity by constructing integrated functional brain network models across individuals, enabling robust cross-subject decoding performance. We validated the effectiveness of MIBRAIN by decoding neural representations underlying both audible and silent articulation of Mandarin Chinese phonemes, with a particular focus on initial consonants. Alongside the 21 distinct initial consonants, our analysis included two simple phonetic finals, 'i' and 'u', due to their frequent use as semi-consonants or syllable-initial elements preceding other vowels. Notably, compound vowel finals, such as 'ou' and 'ao', were omitted from our analysis. These compound vowels pose significant decoding challenges, even within single-subject paradigms, because their articulation dynamically transitions between distinct phonetic states, complicating neural decoding. For example, articulating the compound vowel 'ao' involves a sequential transition from 'a' toward 'o', resulting in intermediate phonetic states that closely resemble those produced during the reversed articulation sequence 'oa'. Such subtle and dynamic transitions yield highly similar articulatory patterns that are challenging to differentiate. Therefore, as a proof of concept, our current investigation prioritized decoding initial consonant articulations in combination with the two selected simple finals.

Systematic offline evaluations involving audible and silent speech decoding tasks on a cohort of 11 participants demonstrated MIBRAIN's superior decoding performance compared to existing state-of-the-art methods and baseline benchmarks. Further analysis of confusion matrices organized according to the place of articulation provided insights into why integrating neural data across multiple participants enhances decoding accuracy. Specifically, MIBRAIN is able to identify and leverage nuanced articulatory features, effectively distinguishing initial phonemes that share similar articulatory characteristics. Additionally, MIBRAIN's robust and consistent performance in real-time online decoding underscores its promise for unified neural representation learning in practical real-world applications. Notably, we observed improved decoding performance scaling continuously as data from additional participants were incorporated into the model, highlighting a promising scalability potential of MIBRAIN for future use in larger-scale applications and clinical contexts.

MIBRAIN's capability to impute neural representations for brain regions without direct recordings positions it as a promising approach for modeling whole-brain network. By integrating subject-specific partial neural network from multiple individuals, MIBRAIN effectively constructs a unified, comprehensive cross-subject functional neural network. Importantly, this strategy addresses inherent heterogeneity arising from the variability of electrode implantation sites and individual neurophysiological differences. Our results demonstrate that neural decoding using solely MIBRAIN's imputed neural representations achieves performance exceeding state-of-the-art methods based on direct recordings. This finding underscores MIBRAIN's critical potential applicability to patient populations with severe neurological impairments, for whom recording neural signals in several certain brain regions can be challenging or impossible. For example, current speech neural prostheses rely heavily on electrodes placed primarily within the ventral sensorimotor cortex to decode phoneme articulation. Substantial impairments in this critical region restricts the direct application of conventional decoding approaches. MIBRAIN has the potential to address this limitation by extracting neural features from brain regions situated upstream of the impaired areas within the language network, imputing missing neural representations. Alternatively, MIBRAIN can leverage neural information aggregated across multiple individuals to reconstruct representations in impaired regions. Furthermore, the robust decoding performance on unseen subjects demonstrated by MIBRAIN significantly expands potential applications, particularly for subjects who either possess severe neurological deficits or are unable to undergo prolonged recording sessions. Additionally, MIBRAIN's ability to reliably impute neural activity holds promise for substantially reducing the data-collection burden inherent in conventional clinical and research procedures.

A major challenge in current neuroscientific research is how to capture the shared neural activity pattern across subjects and align this pattern into a common neural representational space. Here, MIBRAIN showed its ability to identify the neural network of speech production in a group level. The most significant contribution was identified in precentral gyrus and postcentral gyrus which have been extensively researched in speech decoding[7,9,29]. More interestingly, we found an early involvement in SFG (500ms before speech onset), making SFG a better speech onset detector than sensorimotor cortex and thus lower the latency in real-time speech decoding[8,15]. In temporal lobe, inferior temporal gyrus, fusiform gyrus and parahippocampus gyrus activate during speech production. These regions have been termed as the basal temporal language area (BTLA) which, as a part of ventral pathway, is proven to be related with multiple language tasks (e.g., naming and reading)[30,31]. Besides these cortical regions, subcortical regions (e.g., hippocampus, amygdala and thalamus) also involve in speech production. It is important to note that we did not perform a similar contribution analysis for silent articulation. Unlike audible trials, silent articulations lack corresponding audio cues, precluding exact identification and alignment of articulation onset. Furthermore, the variability in participants' reaction times to visual prompting cues and their individual pronunciation habits introduces additional complexity, rendering accurate onset alignment infeasible for the silent articulation task. Beyond the isolated regions, MIBRAIN is capable to reveal the interaction between these regions. We found broad collaboration not only within language network, but also between various cognitive networks. These findings underscore the complex and interconnected nature of language processing and demonstrated the potential of our framework to research the neural basis of cognitive functions in a group-level.

Although MIBRAIN effectively generalizes to unseen subjects, our current approach employs participant-specific embeddings that necessitate ensemble-based decoding for predictions on unseen subjects. Developing a unified encoding strategy could enhance the generalization capabilities and computational efficiency of our proposed model. Additionally, our participant cohort consists exclusively of epilepsy patients with intact language functions; future studies should include patients with diverse language disorders to validate and extend our findings. Another limitation is the considerable impact of mispronunciations on decoding accuracy. This issue warrants immediate attention as mispronunciations are prevalent among patients who significantly benefit from speech neuroprostheses. Lastly, while we have demonstrated MIBRAIN's effectiveness specifically for phoneme articulation, future research should investigate its potential as a versatile general-purpose model applicable to various tasks, thereby enhancing its scalability and practical utility.

## Conclusion

Overall, this study presents a novel neural decoding framework called MIBRAIN that achieves robust and unified decoding performance across subjects. By explicitly modeling the functional neural networks underlying cognitive tasks, MIBRAIN enhances decoding accuracy when integrating neural recordings from multiple subjects and exhibits strong generalization capabilities for previously unseen individuals. Furthermore, MIBRAIN serves analytical tools for comprehensive, group-level exploration of collaborative interactions among brain regions and inter-subject neural synchrony. Future research directions will focus on finer-grained modeling of brain networks, explicitly incorporating inter-hemispheric connectivity and detailed functional reconstruction within anatomically defined brain regions.

# Methods

**Study participants**

This study enrolled a total of 11 participants with five females and six males. The participants were undergoing epilepsy treatment with sEEG electrodes surgically implanted to monitor epileptic seizures for 1-2 weeks at the First and Second Affiliated Hospital of Zhejiang University School of Medicine. We provide detailed participant characteristics, including sex, age, handedness, number of electrodes, and number of selected channels in Tab. S1. All participants were native Mandarin Chinese speakers. Prior to data collection, all patients were informed of study procedure and signed informed consent form to participate. The study was approved by the Ethics Committee of the First and Second Affiliated Hospital of Zhejiang University School of Medicine.

**Task of phoneme articulation decoding**

As a proof of concept, we decoded phonemes from intracranial neural signals recorded while participants read Mandarin Chinese characters. The reading material comprised 407 unique characters encompassing all possible Mandarin Chinese phonetic syllables, thereby mimicking natural speech patterns and enhancing the generalizability of decoding results. A detailed list of these 407 characters and their corresponding syllables is provided in Tab. S2–S5. During data collection (Fig. 1a, upper panel), participants read each of these 407 characters three to five times (3–5 trials per character), either audibly or silently, depending on individual enrollment durations. Each reading trial allotted 0.8–1.2 seconds of preparation, followed by 2 seconds for articulation, and 0.5–1.0 seconds of rest. A complete set of 407 characters constituted one session, with each session containing either audible or silent articulation trials presented in random order. During the online decoding experiment test, participants were comfortably positioned on their beds, viewing a laptop that displayed the target characters alongside a visual countdown cue indicating when articulation should commence. To mitigate fatigue, we randomly selected three characters starting with each initial phoneme, resulting in a total test set of 69 characters. During the free practice period, the real-time decoding results were promptly displayed to participants after processing for their reference: a correct prediction was indicated with the word "Correct" appearing onscreen whenever at least one model yielded the correct prediction, whereas incorrect predictions showed the top three predicted initials ranked by confidence scores. Following this practice period, participants completed four additional decoding sessions—two sessions for each articulation mode. Prior to formal testing, participants completed an introductory session to familiarize themselves with the system and a brief warm-up task involving 10 randomly selected characters.

**Anatomical localization and visualization of electrodes**

We utilized SPM12 for co-registering each participant's preoperative T1 MRI with their postoperative CT (which includes electrode locations). Subsequently, we employed the FreeSurfer neuroimaging analysis software[32] to reconstruct the pial surface and determine the anatomical structure where each contact (channel) is located. The standard Montreal Neurological Institute template brain[33] was used for visualization in Fig. 1c.

**Recording and preprocessing of neural signal**

The number and anatomical placement of electrodes were determined exclusively by clinical requirements for epilepsy treatment. Neural signals for participants 1–4 were recorded using a Neurofax EEG-1200 system (Nihon Kohden Corporation, Japan) at a sampling rate of 2,000 Hz, whereas signals for participants 5–11 were acquired using a Natus Quantum LTM Amplifier at 2,048 Hz. The raw recordings were low-pass filtered at 200 Hz to prevent aliasing and notch filtered at 50 Hz and 100 Hz to remove line noise. Channels exhibiting poor signal quality—either a low signal-to-noise ratio or notable abnormalities observed during manual inspection—were removed, and the remaining data were downsampled to 512 Hz. To identify channels linked to phoneme articulation, we computed the power spectral density (PSD) for each cortical channel (excluding those in white matter) using Welch's method and compared mean PSD values in the 4–150 Hz band between articulation and resting conditions via paired t-tests. We applied Benjamini-Hochberg corrections to control the false discovery rate[34], retaining channels that showed significant differences (corrected P < 0.05). Furthermore, electrodes located within the visual cortex (e.g., in the occipital lobe) were excluded to prevent potential interference from visual processing. The final set of selected channel locations across participants is illustrated in Fig. 2c, with detailed channel positions for individual participants across corresponding brain regions provided in Fig. S1.

**Recording and preprocessing of audio signal**

Microphone recordings were collected simultaneously with neural data using a directional microphone (Audio-Technica) sampled at 44.1 kHz, during both the audible and silent articulation tasks. To avoid contamination, audio and neural signals were separately recorded and stored. To ensure data quality, we quantified correlations between neural and audio signals using randomly selected 10-minute intervals, confirming that audio recordings did not contaminate neural data (Fig. S2). In addition, all audio recordings underwent expert linguistic review to identify and exclude samples containing pronunciation errors; voiced samples were additionally removed from the silent articulation dataset. Neural data corresponding to these excluded audio samples were also discarded from subsequent analyses. Three phonetics experts independently labeled phoneme articulation for each audio segment, and these labels were temporally aligned with neural recordings using PRAAT software[34], marking the articulation onset timestamps accordingly.

**Whole functional network prototyping**

Given a neurological decoding task or a specific cognitive state, we first define a comprehensive set of brain regions, denoted as $\mathcal{R}^c$, with cardinality $N^c$, which are considered functionally related to the task or state by a neurologist. We collect intracranial recordings from m subjects, represented as $S = \{s_i\}_{i=1}^m$, each providing neural responses to the same neural stimuli. We assume significant differences exist among the m sets of recordings $\{X_i\}_{i=1}^m$, primarily due to variations in electrode configurations across subjects. For each subject $s_i$, we assume that electrodes are implanted in brain regions denoted as $R_i$, where the number of regions of interest is $N_i^p = |\mathcal{R}^c \cap \mathcal{R}_i|$. Let $X_i \in \mathbb{R}^{M_i \times C_i \times T}$ represent the set of recordings collected from subject $i$, where $M_i$ denotes the number of data samples, $C_i$ is the number of channels, and $T$ is the segment length (which remains the same across subjects). $C_{i,j}$ denotes the number of channels in the $j^{th}$ brain region of the subject $s_i$ and the total number of channels for subject $i$ is $C_i = \sum_{j=1}^{N_i^p} C_{i,j}$. MIBRAIN first maps the neural recordings $X_i$ for each subject $i$ into brain regional neurological features, denoted as $z_i = \mathcal{E}_i(x; \theta_i) \in \mathbb{R}^{N_i^p \times T' \times d}$, using a set of neuro-embedding encoders $\{\mathcal{E}_i(\cdot; \theta_i)\}_{i=1}^m$, parameterized by network parameters $\{\theta_i\}_{i=1}^m$. $T'$ denotes the temporal feature length and $d$ is the feature dimension in the projected neural space. We utilize brain-region-wise filter banks for neural encoding. Specifically, for the brain region $r$ of subject $i$, we apply 1-D convolution filters $K_i^r(\cdot)$ that take the neural recordings of the region $r$ with an input channel count of $C_i^r$ and produce a $d$-dimensional neural token that captures the neural activity within that region.

Next, we define a set of learnable brain region prototype tokens, denoted as $\mathcal{P} = \{p_k \in \mathbb{R}^{T' \times d}\}_{k=1}^{N^c}$. For each subject $i$, we concatenate the partial neural tokens $z_i$ with the learnable brain region prototype tokens $z^p$ corresponding to the brain regions $\mathcal{R}^p = \mathcal{R}^c \setminus \mathcal{R}_i$, which have no electrode implantation. This concatenation forms a complete region network representation:

$$z_i^c = \text{reorder}(\text{concat}(z_i, z^p)) \in \mathbb{R}^{N^c \times T' \times d}, \tag{1}$$

where $\text{concat}(\cdot)$ represents the channel-wise concatenation operation, and $\text{reorder}(\cdot)$ reorders the brain region tokens within the brain region dimension based on a predefined order to maintain consistency across subjects.

During the whole functional network prototyping phase, our objective is to learn a set of brain region prototype tokens that capture generic neural features shared across subjects. To achieve this, we introduce a masked autoencoding-based pretext task. The underlying intuition is that the signal can be effectively reconstructed by incorporating information from both unmasked and functionally related brain regions within the same subject or by leveraging unmasked regions from other subjects with similar functional properties. We randomly mask a subset of the region tokens in $z_i$ using a masking ratio $r$, drawn from a uniform distribution. During training, the masked tokens are replaced with the corresponding neural tokens from the brain region prototype set $\mathcal{P}$. The masked and concatenated complete neural tokens, denoted as $\hat{z}_i^c$, are then passed through a temporal CNN encoder $\mathcal{E}_T(\cdot)$ for further feature extraction, producing an output $\bar{z}_i^c \in \mathbb{R}^{N \times T'' \times D}$, where $D$ is the token dimension and $T''$ represents the temporal length. Subsequently, we apply a set of lightweight reconstructors $\{R_i(\cdot; \phi_i)\}_{i=1}^m$ to map $\bar{z}_i^c$ back to the original intracranial recording space, denoted as $\widehat{\mathcal{X}}_i$. The training objective is defined as the mean squared error (MSE) between the original input recordings $\mathcal{X}_i$ and the reconstructed signal $\widehat{\mathcal{X}}_i$:

$$\mathcal{L}^{\text{hbp}} = \sum_{i=1}^m \| \mathcal{X}_i - \widehat{\mathcal{X}}_i \|_2^2. \tag{2}$$

**Neural Decoding**

During the neural decoding phase, we discard the reconstructors and reuse the neuroembedding encoders $\{\mathcal{E}_i\}_{i=1}^m$, the temporal encoder $\mathcal{E}_T(\cdot)$ and the set of learnable brain region prototype tokens $\mathcal{P}$ initializing the corresponding weights with those pre-trained during the whole brain prototyping stage. Again, the neurological features $z_i$ are concatenated with the brain region prototype tokens $z^p$ and then passed through the temporal encoder $\mathcal{E}_T(\cdot)$ with the output $\bar{z}_i^c$. We then design a region attention encoder $\mathcal{E}_{RA}(\cdot)$ that explores the collaborative relationships between different brain regions by calculating the functionality similarities of different brain regions. We merge the similar brain regions gradually in the region attention encoder as $z_i^R = \mathcal{E}_{RA}(\bar{z}_i^c) \in \mathbb{R}^{N^m \times T'' \times D}$, where the number of merged brain regions $N^m < N^c$. In each block, the Multi-head Self-attention (MHSA) and feed-forward network (FFN) could be calculated, where the MHSA of each head is defined as:

$$\text{MHSA}^{l,h}(\bar{z}_i^c) = \sum_{b \leq a} S_{a,b}^{l,h}(\bar{z}_{i,a,b}^c W_{OV}^{l,h}), \tag{3}$$

$$S_{a,b}^{l,h} = \text{softmax}\left(\frac{(\bar{z}_{i,a,b}^c W_Q^{l,h})(\bar{z}_{i,a,b}^c W_K^{l,h})^T}{\sqrt{D_h}}\right), \tag{4}$$

where $a, b \in [1, N^c]$ are the indices of tokens of brain regions, the $l$ and $h$ denote the $l^{th}$ block with the $h^{th}$ attention head, $D_h$ denotes the embedding dimension of the attention query and key $W_Q^{l,h}, W_K^{l,h} \in \mathbb{R}^{D \times D_h}, W_{OV}^{l,h} \in \mathbb{R}^{D \times D}$ represents the attention value and output projection. We can get a similarity matrix for token merging in each region as the mean of attention map, $S_i \in \mathbb{R}^{N^c \times N^c}$. Then, the features of merged regions can be produced by two steps, grouping and merging, formulated as follows:

$$z_i^R = \text{Merge}(\bar{z}_i^c, \text{Group}(S_i)), \tag{5}$$

where the output of grouping $A_i = \text{Group}(S_i) \in \{0,1\}^{N^c \times N^c}$ denotes the adjacent matrix with $N_i^m$ connected components. Following ToMe[35], we adopt Bipartite Soft Matching (BSM) as the grouping operator $\text{Group}(\cdot)$, which divides the input tokens into two disjoint sets $\mathbb{G}_s$ and $\mathbb{G}_t$ and constructs a bipartite graph for a sparse adjacency matrix $A \in \{0,1\}^{|\mathbb{G}_s| \times |\mathbb{G}_t|}$. Then, the merging operator $\text{Merge}(\cdot, \cdot)$ is implemented by pooling that combines the connected features.

**Framework architectures**

We implemented the neuro-embedding encoders using two sequential one-dimensional convolutional blocks. Each block comprised a one-dimensional convolutional layer followed by batch normalization, with leaky rectified linear units as the activation function. A dropout rate of 0.1 was applied for regularization. The first and second convolutional blocks used kernel sizes of 7 and 5 and strides of 4 and 2, respectively. The resulting features were projected to a fixed embedding dimension ($d = 32$), and the temporal dimension was downsampled by a factor of 8, i.e., $T' = \frac{1}{8}T$. Subsequently, the temporal CNN encoder utilized four group-wise convolutional blocks, each containing a one-dimensional temporal convolutional layer with a kernel size of 3, stride of 2, and rectified linear unit activation. This stage further reduced the temporal dimension by a total factor of 16 (i.e., $T'' = \frac{1}{16}T'$). The hidden dimensions for these four blocks were set to $1d, 2d, 4d, 4d$, respectively, with the output token dimension $D = 4d$. Finally, the region attention encoder incorporated two MHSA blocks, each with an FFN ratio of 4.

**Experimental setup**

The experiments were conducted using PyTorch[36] on NVIDIA H800 GPUs. For offline model performance evaluations on the prediction accuracy with missing brain regions and the performance of the multi-subject model, intracranial recordings from multiple trials were combined and divided into training and evaluation sets. During training, the training data were integrated from multiple subjects, while the evaluation performance was calculated on a per-subject basis. Specifically, for each subject, we randomly selected seven samples from each initial class, yielding a total of 161 test samples for evaluation, with the remaining recordings used for training and validation. In the offline evaluation of the model performance on a left-out unseen subject, all data from subjects other than the left-out subject were used for model training, while all data from the left-out subject were used for evaluation. To evaluate the scaling effect based on the number of subjects, experiments were conducted on two subjects who enrolled in the online decoding task. For each test subject, data from previously enrolled subjects were incorporated in sequential order, using all data from those subjects as training data and the recordings from the test subject as the test set. For online phoneme decoding evaluations, data collected from all subjects were used for model training (MIBRAIN-multi-sub).

During the whole functional network prototyping stage, we use AdamW as the optimizer with a base learning rate of $1 \times 10^{-5}$. Cosine annealing decay is applied for learning rate scheduling. We set momentum factors $\beta_1, \beta_2 = 0.9, 0.999$, along with a weight decay of 0.005. The batch size is set to 16 times the number of subjects involved in the model training. A fixed number of 1000 epochs is used for training. For the neural decoding stage, we again use AdamW as the optimizer, but with a higher base learning rate of $3\times 10^{-4}$. Cosine annealing decay is also adopted for learning rate scheduling, and we maintain the same momentum factors as in the whole brain prototyping stage, with a weight decay of 0.001. In supervised neural decoding, the model is trained for a fixed number of 200 epochs. For the training of baseline approaches, we adhere to the experimental setups provided by the respective authors.

**Neural Decoding on Unseen Subject**

Given the trained neuro-embedding encoders $\{\mathcal{E}_i\}_{i=1}^m$, the temporal encoder $\mathcal{E}_T(\cdot)$, the set of learnable brain region prototype tokens $\mathcal{P}$, the region attention encoder $\mathcal{E}_{RA}(\cdot)$, and the neural prediction heads trained using data from the existing m subjects, we perform neural decoding on a $(m+1)^{th}$ subject via a majority voting mechanism. The neural tokens produced by the neuro-embedding encoders are concatenated with the brain region prototype tokens to form a complete-region brain representation, providing a unified representation for subsequent decoding. So, we seek to embed the neurological recordings from the $(m+1)^{th}$ subject with existing neuro-embedding encoders $\{\mathcal{E}_i\}_{i=1}^m$ by maximizing the homogeneity among the different subjects in a pairwise fashion. We calculate the pairwise channel similarity of every brain region $r \in \mathcal{R}_{m+1} \cap \mathcal{R}_i$. Given brain recordings $X_i^r$ of brain region $r$, we define a channel similarity matrix $S^r \in \mathbb{R}^{C_{m+1} \times C_i}$:

$$S^r(c_1, c_2) = \frac{\langle \bar{X}_{i,c_1}^r, \bar{X}_{m+1,c_2}^r \rangle}{\| \bar{X}_{i,c_1}^r \| \| \bar{X}_{m+1,c_2}^r \|}, c_1 = 1, \dots, C_1, c_2 = 1, \dots, C_2, \quad (7)$$

where $\bar{X}_{i,c}^r = \frac{1}{M_i} \sum_{n=1}^{M_1} X_{i,n,c,t}^r$, $c = 1, \dots, C_i$. We then find the most similar channel $c_2$ of subject $m+1$ to the channel $c_1$ of subject $i$:

$$c_2^*(c_1) = \mathrm{argmax}_{c_2 \in \{1,\dots,C_{m+1}\}} S^r(c_1, c_2), c_1 = 1, \dots, C_i. \quad (8)$$

We then use the convolutional filters $K_i^r(\cdot)$ bank of subject $i$ that of neuro-embedding encoders $E_i$ encodes brain region $r$ to embed the brain recordings of subject $m+1$. In particular, we create re-assembled brain recordings of subject $m+1$ by reorganizing the channels:

$$\hat{X}_{m+1}^r = \begin{cases} \text{concat}(X_{m+1,c_2^*(c_1)}^r), (c_1 = 1, \ldots, C_i) & C_i \leq C_{m+1} \\ \text{zeropadding}(\text{concat}(X_{m+1,c_2^*(c_1)}^r)), (c_1 = 1, \ldots, C_i) & C_i > C_{m+1} \end{cases}, \quad (9)$$

where $\text{pad}(\cdot)$ is padding $C_i - C_{m+1}$ channels of zero vectors $\mathbf{0} \in \mathbb{R}^{M_{m+1} \times 1 \times T}$. We then perform the same operation to every brain region $r \in \mathcal{R}_{m+1} \cap \mathcal{R}_i$ to get neural tokens $z_{m+1}(i)$ of subject $m+1$ with respect to subject $i$. We then create the complete-region brain representation $z_{m+1(j)}^c$ by adding the learnable region prototypes of brain regions in $\mathcal{R}^c \setminus (\mathcal{R}_{m+1} \cap \mathcal{R}_i)$. By passing $z_{m+1(j)}^c$ to $\mathcal{E}_T(\cdot)$, $\mathcal{E}_{RA}(\cdot)$ neural prediction head, we get a prediction result $\hat{y}_{m+1}(i)$. We repeat the same process and get the prediction results with respect to all m subjects. Finally, the prediction result is given by taking the majority voting on results $\{\hat{y}_{m+1(i)}\}_{i=1}^m$.

**Calculation of brain region contribution**

Given the intracranial recordings $X_i \in \mathbb{R}^{M_i \times C_i \times T}$ from subject $i$, we define $\hat{Y}_i \in \mathbb{R}^{M_i \times G}$ to be the corresponding one-hot label vector of the underlying neurological decoding task with $G$ classes. Recall that $z_i^c \in \mathbb{R}^{N \times T' \times d}$ is the complete-region brain representation computed from $X_i$ with the neuro-embedding encoders and concatenated with the brain region prototype tokens. We calculate the region contribution score $\zeta(r, t)$ based on the class activation maps [37] as:

$$\zeta(r, t) = \frac{1}{m \times d} \sum_{i=1}^{m} \sum_{l=1}^{d} \sum_{j=1}^{M_i} \frac{1}{M_i} \cdot \text{ReLU}\left(\frac{\partial \hat{Y}_{i,j}}{\partial z_{i,j,r,t,l}}\right). \quad (6)$$

Here, $\zeta(r, t)$ represents the contribution score of brain region $r$ at the timestamp $t$, averaged over all data samples across all subjects.

**Calculation of Brain Region Collaboration**

For each subject $S_i$, we define their neural recordings as $X_i \in \mathbb{R}^{M_i \times C_i \times T_i}$, where $M_i$ denotes the number of recording clips, $C_i$ represents the number of channels, and $T_i$ indicates the temporal length of each recording. To characterize the dynamic collaborative relationships between brain regions, we create sparse adjacency matrices

$A_i \in \{0,1\}^{M_i \times T_i'' \times G_i \times N^c}$. For each recording clip $m$ at time point $t$, the source matrix $A_{i,m,t} \in \{0,1\}^{G_i \times N^c}$ comprises column vectors $A_{i,m,t,*,r} \in \{0,1\}^{N^c}$, $0 \leq r < N^c$. Each column vector is a one-hot encoded representation where $A_{i,m,t,g,r} = 1$, $0 \leq g < G_i$ indicates that brain region $r$ belongs to group $g$. Correspondingly, each row vector $A_{i,m,t,g} \in \{0,1\}^{N^c}$ defines a set of brain regions $\mathcal{R}_{i,m,t,q}^s = \{r \in \mathcal{R}_i \mid A_{i,m,t,g,r} = 1\}$ that form a functional group.

To facilitate the analysis of brain region relationships, we construct a set of adjacent matrices that capture the grouping patterns between regions. For each subject $s_i$ recording clip $m$, and time point $t$, we define a collaboration matrix $C_{i,m,t} = A_{i,m,t}^T S_{i,m,t} \in \{0,1\}^{N^c \times N^c}$, where $C_{i,m,t,j,k} = 1$ indicates that brain regions $j$ and $k$ are assigned to the same group at time $t$ in recording $m$. This transformation allows us to directly represent pairwise relationships between brain regions, making it easier to analyze co-activation patterns and functional connectivity.

By normalizing over all recording clips, we can obtain a grouping frequency matrix $C_i^S = \frac{\sum_{m=1}^{M_i} A_{i,m,t}}{M_i} \in \mathbb{R}^{T_i'' \times N^c \times N^c}$, where each element $C_{i,t,j,k}^S$ represents the frequency of brain regions $j$ and $k$ being grouped together at time $t$ across all recording clips. This normalized matrix provides a quantitative measure of the temporal consistency in region groupings and reveals persistent functional relationships between brain areas. Higher values in $C_i^S$ indicate more frequent grouping of regions, suggesting stronger functional associations or consistent collaborative patterns in neural activity. To identify common patterns of brain region interactions across subjects, we construct an aggregated grouping collaboration matrix by calculating the average.

## Acknowledgements

The project was funded by STI2030-Major Project (2022ZD0208805), National Natural Science Foundation of China (Grant No. 623B2085), and "Pioneer" and

## Author contributions

Di Wu: Conceptualization, Methodology, Software, Visualization, Funding acquisition, Writing - Original Draft. Linghao Bu: Conceptualization, Investigation, Writing - Review & Editing. Yifei Jia: Software, Visualization, Writing - Original Draft. Lu Cao: Data Curation, Project administration. Siyuan Li: Methodology, Software, Writing - Original Draft. Siyu Chen: Software, Data Curation. Yueqian Zhou: Data Curation, Visualization. Sheng Fan: Software, Data Curation. Wenjie Ren: Visualization. Dengchang Wu: Data Curation. Kang Wang: Writing - Review & Editing. Yue Zhang: Project administration, Supervision, Writing - Review & Editing. Yuehui Ma: Project administration, Supervision, Writing - Review & Editing. Jie Yang: Project administration, Funding acquisition, Supervision, Writing - Review & Editing. Mohamad Sawan: Resources, Funding acquisition, Supervision, Writing - Review & Editing.

## Data availability

In accordance with clinical ethical guidelines, the raw data involved in this study, including audio recordings and neurophysiological signals, cannot be publicly distributed. However, electrophysiological data may be made available upon reasonable request directed to the corresponding author. To protect participant anonymity, any potentially identifiable information will be removed from shared datasets, and all provided data must remain confidential and not be redistributed. Source data underlying figures are provided with this paper.

## Competing interests

The authors declare that there is no conflict of interest regarding the publication of this article.

# Supplementary Materials

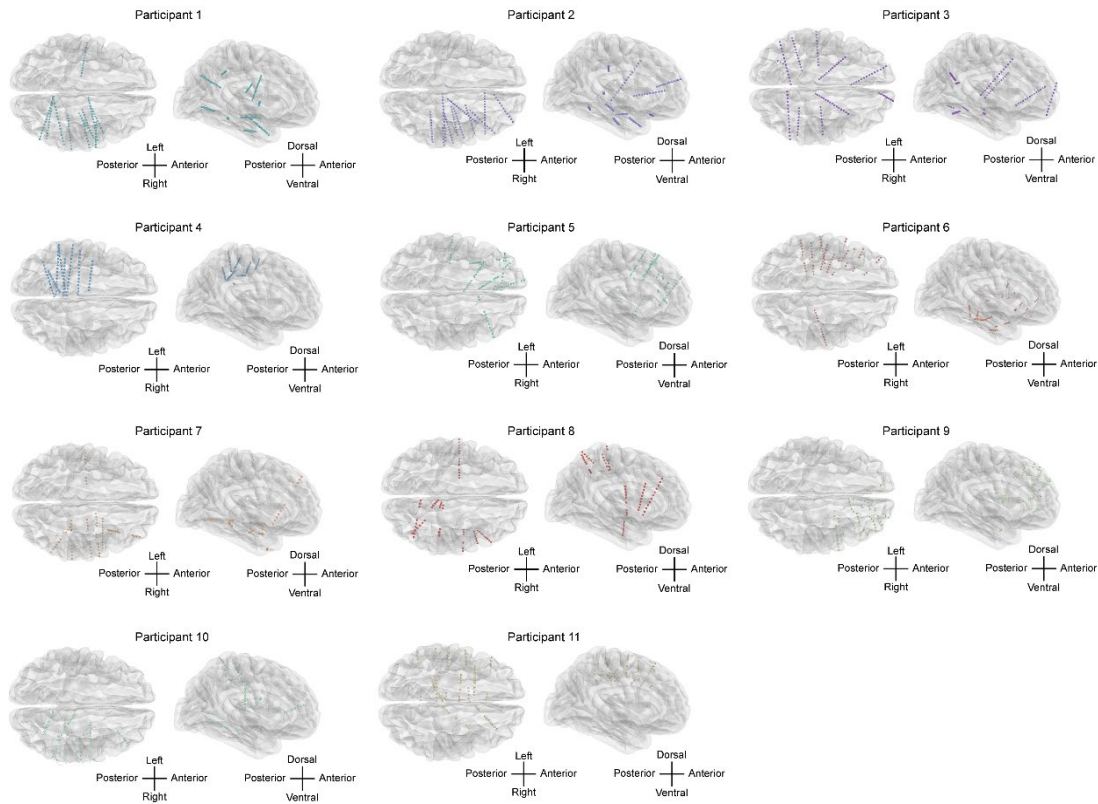

**Fig. S1.** | Implantation sites of selected electrodes for each participant are illustrated on the standard Montreal Neurological Institute (MNI) template brain. Electrodes from each participant are depicted using distinct colors to clearly differentiate implantation sites. Anatomical orientation indicated by directional labels ensures clarity.

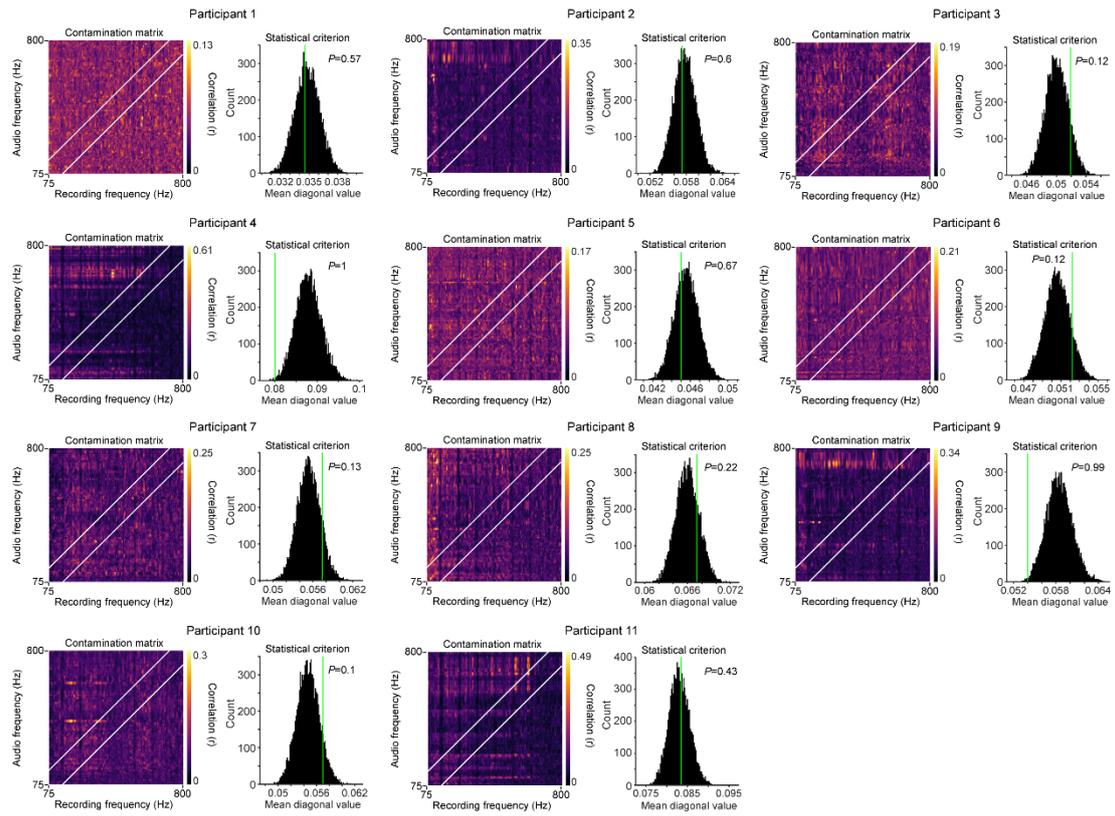

**Fig. S2. | Objective assessment of acoustic contamination in neural recordings for all participants.** Left panels: Heatmaps depicting contamination matrices computed from neural data of each participant, with brighter colors indicating higher correlations between neural signals and acoustic inputs. Right panels: Statistical evaluation of contamination strength, comparing the observed average diagonal values in each participant's contamination matrix against the null distributions obtained from 10,000 randomly shuffled matrices.

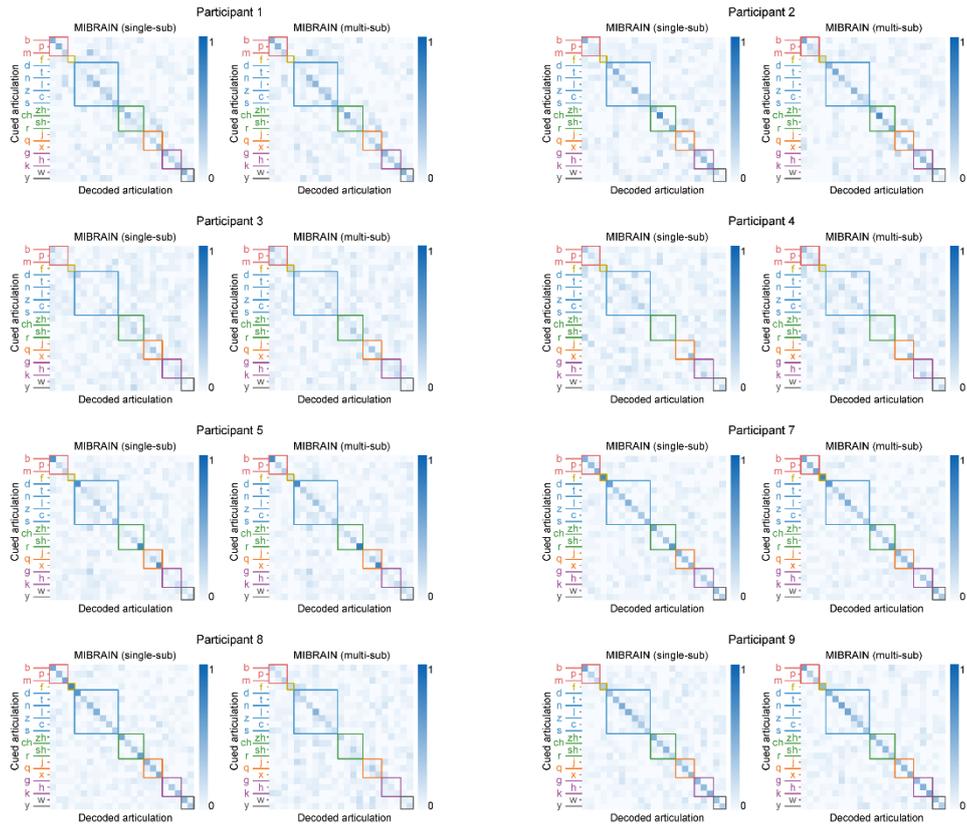

**Fig. S3.** | Audible phoneme decoding accuracy displayed by consonant initials grouped according to place of articulation, comparing MIBRAIN (single-sub) and MIBRAIN (multi-sub) performance for additional participants.

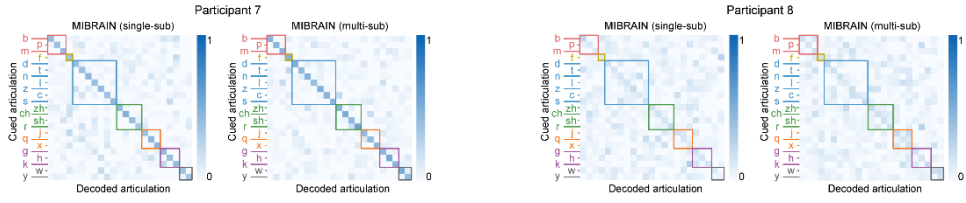

**Fig. S4.** | Offline silent phoneme decoding accuracy by consonant initials using both MIBRAIN variants (single-sub and multi-sub) for participants 7 and 8.

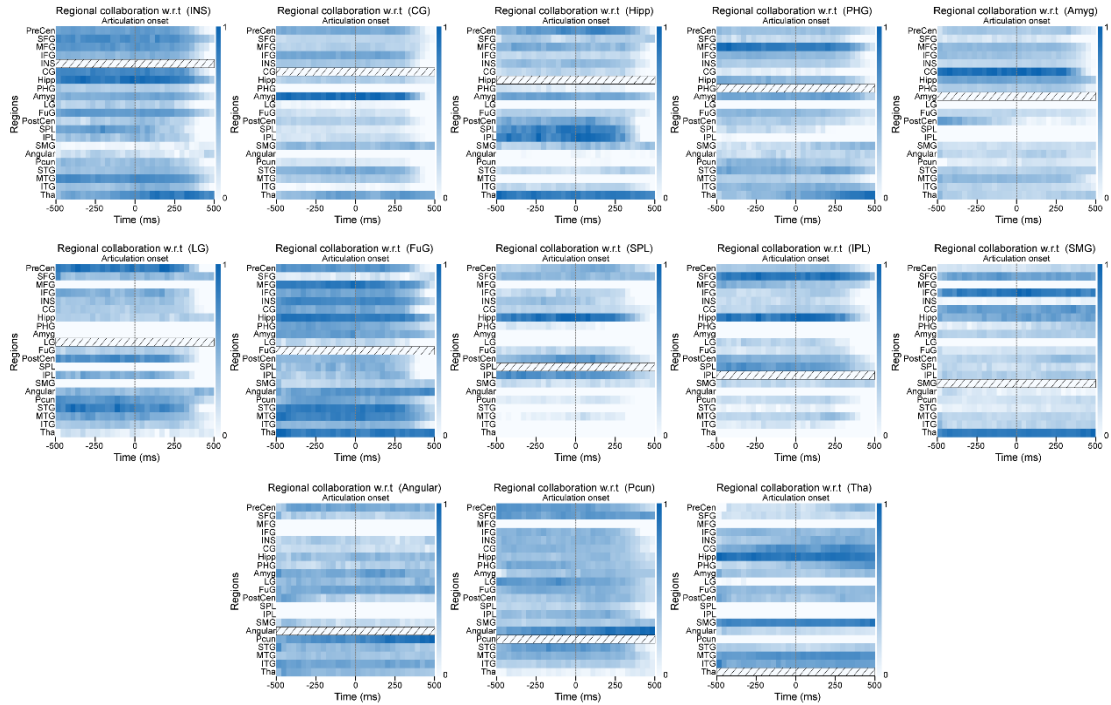

**Fig. S5.** | Functional collaboration of brain regions with respect to a target region (indicated by rectangles filled with diagonal lines). Brain region abbreviations are as follows: INS, insula; CG, cingulate gyrus; Hipp, hippocampus; Amyg, amygdala; LG, lingual gyrus; FuG, fusiform gyrus; SPL, superior parietal lobule; IPL, inferior parietal lobe; SMG, supramarginal gyrus; Angular, angular gyrus; Pcun, precuneus; Tha, thalamus.

**Table S1.** Participants statistics including sex, age, handedness and number of implanted electrodes enrolled in this study.

| Participant | Sex | Age | Handedness | Number of electrodes | Selected channel count |
|---|---|---|---|---|---|
| 1 | Male | 20 | Right-handed | 12 | 63 |
| 2 | Male | 25 | Right-handed | 8 | 36 |
| 3 | Female | 19 | Right-handed | 11 | 30 |
| 4 | Female | 19 | Right-handed | 13 | 66 |
| 5 | Male | 14 | Right-handed | 9 | 73 |
| 6 | Female | 17 | Left-handed | 11 | 78 |
| 7 | Female | 15 | Right-handed | 11 | 74 |
| 8 | Male | 15 | Right-handed | 12 | 68 |
| 9 | Female | 26 | Right-handed | 12 | 61 |
| 10 | Male | 35 | Right-handed | 12 | 88 |
| 11 | Male | 22 | Right-handed | 10 | 84 |

**Table S2.** The first set of 407 Mandarin Chinese characters used as reading material, organized by syllables according to their initial phonemes.

| Syllables | Characters | Syllables | Characters | Syllables | Characters |
|---|---|---|---|---|---|
| ā | 阿 | chě | 扯 | diào | 吊 |
| ài | 爱 | chèn | 趁 | dié | 叠 |
| ān | 安 | chéng | 城 | dǐng | 顶 |
| áng | 昂 | chī | 痴 | diū | 丢 |
| ǎo | 袄 | chóng | 虫 | dōng | 东 |
| bá | 拔 | chóu | 愁 | dòu | 豆 |
| bái | 白 | chū | 初 | dú | 读 |
| bǎn | 板 | chuā | 欻 | duǎn | 短 |
| bāng | 帮 | chuāi | 揣 | duì | 对 |
| bǎo | 保 | chuān | 穿 | dūn | 蹲 |
| bēi | 杯 | chuáng | 床 | duó | 夺 |
| běn | 本 | chuī | 吹 | é | 鹅 |
| bēng | 崩 | chūn | 春 | ēn | 恩 |
| bí | 鼻 | chuō | 戳 | ēng | 鞥 |
| biān | 编 | cì | 次 | ěr | 耳 |
| biāo | 标 | cōng | 葱 | fá | 罚 |
| biě | 瘪 | còu | 凑 | fǎn | 反 |
| bīn | 宾 | cū | 粗 | fāng | 方 |
| bìng | 病 | cuàn | 窜 | féi | 肥 |
| bó | 伯 | cuì | 脆 | fěn | 粉 |
| bǔ | 补 | cún | 存 | fēng | 风 |
| cā | 擦 | cuō | 搓 | fú | 福 |
| cái | 财 | dǎ | 打 | fǒu | 否 |
| cán | 残 | dài | 带 | fù | 付 |
| cāng | 仓 | dǎn | 胆 | gà | 尬 |
| cáo | 槽 | dǎng | 党 | gài | 盖 |
| cè | 测 | dào | 到 | gǎn | 敢 |
| cān | 参 | dé | 德 | gāng | 缸 |
| céng | 层 | děi | 得 | gào | 告 |
| chá | 茶 | dèn | 扽 | gē | 割 |
| chái | 柴 | děng | 等 | gěi | 给 |
| chǎn | 产 | dǐ | 底 | gēn | 根 |
| chàng | 唱 | diē | 爹 | gēng | 耕 |
| chāo | 超 | diàn | 电 | gòng | 共 |

**Table S3.** The second set of 407 Mandarin Chinese characters used as reading material, organized by syllables according to their initial phonemes.

| Syllables | Characters | Syllables | Characters | Syllables | Characters |
|---|---|---|---|---|---|
| gǒu | 狗 | jīn | 金 | lěng | 冷 |
| gū | 估 | jìng | 镜 | lì | 力 |
| guā | 瓜 | jiǒng | 窘 | liǎ | 俩 |
| guài | 怪 | jiǔ | 酒 | lián | 连 |
| guān | 关 | jǔ | 举 | liáng | 良 |
| guǎng | 广 | juān | 捐 | liào | 料 |
| guì | 贵 | jué | 决 | liè | 列 |
| gǔn | 滚 | jùn | 俊 | lín | 林 |
| guǒ | 裹 | kǎ | 卡 | lǐng | 领 |
| hā | 哈 | kāi | 开 | liǔ | 柳 |
| hái | 孩 | kǎn | 砍 | lóng | 龙 |
| hán | 寒 | kāng | 糠 | lóu | 楼 |
| háng | 航 | kào | 靠 | lù | 路 |
| hào | 耗 | kē | 科 | luǎn | 卵 |
| hé | 河 | kè | 克 | lún | 轮 |
| hēi | 黑 | kěn | 肯 | luó | 罗 |
| hèn | 恨 | kēng | 坑 | lü | 滤 |
| héng | 恒 | kǒng | 孔 | lüè | 略 |
| hōng | 烘 | kǒu | 口 | mǎ | 马 |
| hǒu | 吼 | kù | 库 | mǎi | 买 |
| hǔ | 虎 | kuā | 夸 | màn | 慢 |
| huà | 画 | kuài | 快 | máng | 忙 |
| huài | 坏 | kuǎn | 款 | máo | 毛 |
| huàn | 换 | kuáng | 狂 | měi | 美 |
| huāng | 慌 | kuī | 亏 | mén | 门 |
| huǐ | 悔 | kǔn | 捆 | měng | 猛 |
| hūn | 昏 | kuò | 阔 | mǐ | 米 |
| huò | 货 | lā | 拉 | miàn | 面 |
| jī | 机 | lái | 来 | miáo | 苗 |
| jià | 价 | lǎn | 懒 | miè | 灭 |
| jiān | 尖 | làng | 浪 | mín | 民 |
| jiǎng | 桨 | lǎo | 老 | mìng | 命 |
| jiāo | 交 | lēi | 勒 | miù | 谬 |
| jiě | 姐 | léi | 雷 | mó | 魔 |

**Table S4.** The third set of 407 Mandarin Chinese characters used as reading material, organized by syllables according to their initial phonemes.

| Syllables | Characters | Syllables | Characters | Syllables | Characters |
|---|---|---|---|---|---|
| móu | 谋 | pén | 盆 | rǔ | 乳 |
| mù | 木 | pèng | 碰 | ruá | 挼 |
| ná | 拿 | pí | 皮 | ruǎn | 软 |
| nǎi | 奶 | piān | 偏 | ruì | 锐 |
| nán | 男 | piào | 票 | rùn | 润 |
| náng | 囊 | piē | 瞥 | ruò | 弱 |
| nào | 闹 | pǐn | 品 | sǎ | 洒 |
| nè | 讷 | píng | 瓶 | sài | 赛 |
| nèi | 内 | pò | 破 | sǎn | 伞 |
| nèn | 嫩 | pōu | 剖 | sǎng | 嗓 |
| néng | 能 | pǔ | 普 | sāo | 骚 |
| ní | 泥 | qí | 骑 | sè | 涩 |
| nián | 年 | qiā | 掐 | sēn | 森 |
| niáng | 娘 | qiǎn | 浅 | sēng | 僧 |
| niào | 尿 | qiāng | 枪 | shā | 沙 |
| niē | 捏 | qiáo | 桥 | shāi | 筛 |
| nín | 您 | qiè | 窃 | shǎn | 闪 |
| níng | 凝 | qín | 琴 | shāng | 伤 |
| niú | 牛 | qíng | 情 | shāo | 烧 |
| nóng | 农 | qióng | 穷 | shē | 赊 |
| nòu | 耨 | qiú | 球 | shuí | 谁 |
| nú | 奴 | qǔ | 取 | shén | 神 |
| nuǎn | 暖 | quàn | 劝 | shéng | 绳 |
| nuó | 挪 | quē | 缺 | shí | 石 |
| nü | 女 | qún | 群 | shǒu | 手 |
| nüè | 虐 | rán | 然 | shū | 书 |
| ó | 哦 | ràng | 让 | shuā | 刷 |
| ǒu | 藕 | ráo | 饶 | shuài | 帅 |
| pá | 爬 | rè | 热 | shuān | 栓 |
| pái | 牌 | rěn | 忍 | shuǎng | 爽 |
| pán | 盘 | rēng | 扔 | shuǐ | 水 |
| páng | 旁 | rì | 日 | shùn | 顺 |
| pāo | 抛 | róng | 容 | shuò | 硕 |
| péi | 赔 | ròu | 肉 | sǐ | 死 |

**Table S5.** The fourth set of 407 Mandarin Chinese characters used as reading material, organized by syllables according to their initial phonemes.

| Syllables | Characters | Syllables | Characters | Syllables | Characters |
|---|---|---|---|---|---|
| sōng | 松 | wǔ | 午 | zǎo | 早 |
| sōu | 搜 | xǐ | 洗 | zé | 则 |
| sū | 酥 | xiá | 霞 | zéi | 贼 |
| suàn | 算 | xiǎn | 险 | zěn | 怎 |
| suì | 岁 | xiàng | 向 | zēng | 增 |
| sǔn | 损 | xiào | 笑 | zhā | 渣 |
| suǒ | 锁 | xiě | 写 | zhài | 债 |
| tǎ | 塔 | xīn | 心 | zhǎn | 展 |
| tái | 抬 | xíng | 形 | zhǎng | 涨 |
| tán | 谈 | xióng | 熊 | zhǎo | 找 |
| tǎng | 躺 | xiū | 修 | zhē | 遮 |
| tǎo | 讨 | xǔ | 许 | zhè | 这 |
| tè | 特 | xuǎn | 选 | zhēn | 针 |
| téng | 藤 | xué | 学 | zhēng | 蒸 |
| tì | 替 | xún | 寻 | zhí | 直 |
| tián | 田 | yá | 芽 | zhǒng | 肿 |
| tiáo | 条 | yān | 烟 | zhōu | 州 |
| tiě | 铁 | yǎng | 养 | zhǔ | 煮 |
| tíng | 停 | yào | 药 | zhuā | 抓 |
| tǒng | 桶 | yě | 野 | zhuāi | 拽 |
| tōu | 偷 | yī | 衣 | zhuān | 砖 |
| tú | 图 | yín | 银 | zhuàng | 撞 |
| tuán | 团 | yīng | 鹰 | zhuī | 追 |
| tuǐ | 腿 | yō | 哟 | zhǔn | 准 |
| tūn | 吞 | yǒng | 永 | zhuō | 捉 |
| tuō | 拖 | yóu | 油 | zì | 字 |
| wā | 挖 | yǔ | 雨 | zǒng | 总 |
| wài | 外 | yuán | 元 | zǒu | 走 |
| wàn | 万 | yuè | 月 | zǔ | 组 |
| wàng | 忘 | yún | 云 | zuān | 钻 |
| wéi | 围 | zá | 杂 | zuì | 醉 |
| wén | 文 | zāi | 栽 | zūn | 尊 |
| wēng | 翁 | zàn | 暂 | zuǒ | 左 |
| wō | 窝 | zàng | 葬 | | |

(2017).